\documentstyle[12pt,epsfig]{article}

\newcommand{\ee}{e^+e^-}
\newcommand{\as}{\alpha_s}

\newcommand{\xpr}{{x^\prime}}

\newcommand{\MSbar}{$\overline{\sml{MS}}$}
\newcommand{\smMSb}{{\overline{\vsm{MS}}}}
\newcommand{\cl}[1]{{\cal #1}}

\newcommand{\sml}[1]{{\mbox{\scriptsize #1}}}
\newcommand{\nml}[1]{{\mbox{#1}}}
\newcommand{\vsm}[1]{{\mbox{\tiny #1}}}
\newcommand{\Sgt}{\tilde{\Sigma}}

\newcommand{\bv}{{\bf b}}
\newcommand{\pp}{{{\bf p}_\perp}}
\newcommand{\kp}{{{\bf k}_\perp}}

\newcommand{\al}{\alpha}
\newcommand{\be}{\beta}
\newcommand{\D}{\Delta}
\newcommand{\de}{\delta}
\newcommand{\e}{\epsilon}
\newcommand{\g}{\gamma}
\renewcommand{\k}{\kappa}

\newcommand{\s}{\sigma}
\newcommand{\Sg}{\Sigma}

\newcommand{\beq}{\begin{equation}}
\newcommand{\eeq}{\end{equation}}
\newcommand{\bea}{\begin{eqnarray}}
\newcommand{\eea}{\end{eqnarray}}
\newcommand{\nln}{\nonumber\\}
\newcommand{\ind}{\hspace{2.0cm}}

\newcommand{\rf}[1]{(\ref{#1})}
\newcommand{\sect}[1]{\section{#1}\setcounter{equation}{0}
               \hspace{\parindent}\hspace{-0.14cm}}
\newcommand{\subsect}[1]{\subsection{#1}
               \hspace{\parindent}\hspace{-0.14cm}}
\renewcommand{\theequation}
           {\arabic{section}.\arabic{equation}}

\newcommand{\ibid}[3]{{\it ibid.} {\bf #1} (#2) #3}
\newcommand{\jhep}[3]{{\it J.~High Energy Phys.} {\bf #1} (#2) #3}
\newcommand{\npb}[3]{{\it Nucl. Phys.} {\bf B #1} (#2) #3}
\newcommand{\epj}[3]{{\it Eur. Phys.~J.} {\bf #1} (#2) #3}
\newcommand{\plb}[3]{{\it Phys. Lett.} {\bf B #1} (#2) #3}
\newcommand{\prep}[3]{{\it Phys. Rep.} {\bf #1} (#2) #3}
\newcommand{\prd}[3]{{\it Phys. Rev.} {\bf D #1} (#2) #3}

\newcommand{\prl}[3]{{\it Phys. Rev. Lett.} {\bf #1} (#2) #3}

\begin{document}
\begin{titlepage}
\begin{flushright}
July 2000 \\
Cavendish-HEP-00/04 \\
BICOCCA-FT-00-11
\end{flushright}              
\vspace*{\fill}
\begin{center}
{\Large \bf Non-perturbative effects in the W and Z transverse momentum
distribution\footnote{Research supported in part by the 
U.K. Particle Physics and Astronomy Research Council.}}
\end{center}
\par \vskip 5mm
\begin{center}
        A.~Guffanti\\
        Dipartimento di Fisica, Universit\`{a} di Milano-Bicocca,\\
        Via Celoria 16, 20133 Milano, Italy.\\
\par \vskip 5mm
        and \\
\par \vskip 5mm
        G.E.~Smye\\
        Cavendish Laboratory, University of Cambridge,\\
        Madingley Road, Cambridge CB3 0HE, U.K.\\
\end{center}
\par \vskip 2mm
\begin{center} {\large \bf Abstract} \end{center}
\begin{quote}
We use the ``dispersive method'' to investigate non-perturbative effects in the transverse momentum distribution of vector bosons produced in $p\bar{p}$ collisions. The assumption of a non-perturbative modification to the running coupling at low scales leads to additional contributions in impact parameter space proportional to $-b^2\log Q^2$ and $-b^2$. Our results, which we expect to be valid provided $\tau$ is not close to 1, are shown to account for a substantial proportion of the total non-perturbative contribution extracted from data.
\end{quote}
\vspace*{\fill}
\end{titlepage}

\sect{Introduction}
The measured transverse momentum distributions of W and Z bosons at the Tevatron and LHC will provide great insights into the structure of hadrons. Indeed some very interesting data has already been released \cite{run1}. On the theoretical side, the application of perturbation theory has led to a full next-to-leading order calculation \cite{nlo} and next-to-leading log resummations \cite{nll,elve,css,fnr}. These are merged to give what is at present our best perturbative prediction for the distribution. There are two different approaches in the literature \cite{nll} to the resummation of these logarithms: either the contributions are resummed directly in transverse momentum space, or a Fourier transform is performed and the calculation developed in impact parameter space. These approaches were compared in \cite{elve} and shown to differ only for NNN-leading logs. In this article we will use the impact parameter formalism, since it is more readily suited to the calculation at hand. Perturbation theory has however proved insufficient to describe experimental data, and it is necessary to include in calculations large non-perturbative effects. Parametrisations of these contributions have been proposed \cite{css}, and values for parameters have been extracted from data \cite{yuan,lbly}; however a full theoretical understanding of these effects has yet to emerge.

One technique that has had success in describing non-perturbative effects over a wide range of QCD observables is the dispersive approach \cite{dmw}. It is related to the renormalon model (see \cite{ben} for a review) but rests on slightly different foundations. The primary assumption is that one can define a universal running strong coupling $\as(k^2)$ that has no singularities in the complex plane except for a branch cut along the negative real axis. In particular there is no Landau pole, which is an unphysical artefact of perturbation theory. The contribution already accounted for in a fixed-order perturbative calculation is then subtracted off, leaving what is termed the `non-perturbative' correction. The assumption of universality of the running coupling, and thus of the non-perturbative parameters derived from it, is most easily tested using data from event shapes in $\ee$ annihilation \cite{eeev} and deep inelastic scattering \cite{disev}: these are particularly large corrections all proportional to the first moment of the non-perturbative modification to the coupling. Studies indicate that universality is approximately realised, both giving support to the model, and prompting potential refinements of some calculations \cite{blois}.

Power corrections to observables in the Drell-Yan process have previously been studied using the renormalon and dispersive approaches in \cite{prev,bebr,mrinal}. In addition, non-perturbative corrections to a calculation involving a next-to-leading log resummation were studied for the first time in the dispersive approach for the $\ee$ energy-energy correlation \cite{eec}. Here we apply the same technique to the next-to-leading log resummation of the W/Z transverse momentum distribution. A simplified overview of the perturbative resummation is presented in section \ref{calc}, while the power corrections to the distribution are calculated in section \ref{powcor}. There then follows in section \ref{result} a discussion of the results.

\begin{figure}[ht]
\begin{center}
\epsfig{file=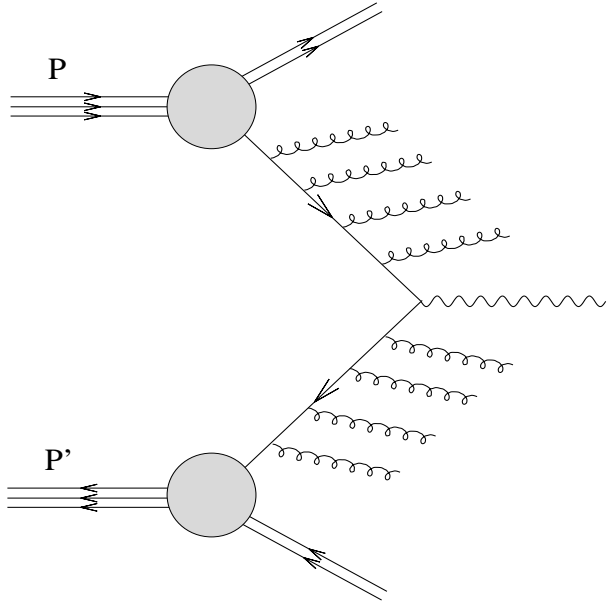, height=8.0cm, width=8.0cm}
\caption{\label{fig}Vector boson production with collinear gluon emission.}
\end{center}
\end{figure}

\sect{Transverse momentum distribution}
\label{calc}
We recall the perturbative prediction for the transverse momentum distribution in two stages: first we consider the emission of a single gluon; and then we resum leading logarithms arising from multiple collinear gluon radiation, figure \ref{fig}. Non-perturbative corrections from the modification to the coupling at low scales can then be added to this perturbative result.

\subsect{Single gluon emission}
Suppose in the collision of two hadrons $P$ and $P^\prime$, a quark and antiquark annihilate into a vector boson. The parton model gives the total cross-section for the process as
\beq
\s(\tau) = \sum_{q,\bar{q}}\int_0^1 dxd\xpr\left[q(x,\mu^2)\bar{q}^\prime(\xpr,\mu^2)+\bar{q}(x,\mu^2)q^\prime(\xpr,\mu^2)\right]\s_{q\bar{q}}(x,x^\prime,\tau,\mu^2)\;,
\eeq
where $q$ and $\bar{q}$ are the annihilating species of quark and antiquark respectively, whose distribution functions in $P$ and $P^\prime$ appear above, $\s_{q\bar{q}}$ is the parton-level cross-section, and the standard Drell-Yan variable is
\beq
\tau = \frac{Q^2}{s}\;,
\eeq
where $Q$ is the boson mass and $\sqrt{s}$ the total energy in centre-of-momentum frame. We have also explicitly introduced the factorisation scale $\mu^2$, which is in principle arbitrary but which (it will be seen below) is naturally chosen for the transverse momentum distribution to be $(b_0/b)^2$.

In order to find the differential $\pp$ distribution $d\s/d^2\pp$, where $\pp$ is the two-component transverse momentum of the produced boson, normal to the incident hadron momenta, we simply insert a factor $\delta^2(\pp+\sum_i\kp_i)$ into the phase-space integral, where here $i$ runs over all emitted gluons. It is convenient to move into impact parameter space, writing
\beq
\delta^2(\pp+\sum_i\kp_i) = \int\frac{d^2\bv}{(2\pi)^2}e^{i\bv\cdot(\pp+\sum_i\kp_i)}\;.
\eeq
Thus we find
\beq
\frac{d\s}{d^2\pp}(\tau) = \int\frac{d^2\bv}{(2\pi)^2}e^{i\bv\cdot\pp}\sum_{q,\bar{q}}\int_0^1 dxd\xpr\left[q(x,\mu^2)\bar{q}^\prime(\xpr,\mu^2)+(\leftrightarrow)\right]\hat{\s}_{q\bar{q}}(x,\xpr,\tau,\mu^2,b)\;,
\eeq
where $\hat{\s}_{q\bar{q}}$ is now the cross-section calculated with an additional factor $e^{i\bv\cdot\kp}$ for each real emitted gluon. Performing the angular integrals then gives
\bea
\label{eqSgm}
\Sigma(\tau,p_\perp^2) &\equiv& \frac{d\s}{dp_\perp^2}(\tau) \\
&=& \frac{1}{2}\int_0^\infty db\,bJ_0(bp_\perp)\sum_{q,\bar{q}}\int_0^1 dxd\xpr\left[q(x,\mu^2)\bar{q}^\prime(\xpr,\mu^2)+(\leftrightarrow)\right]\hat{\s}_{q\bar{q}}(x,\xpr,\tau,\mu^2,b)\;.\nonumber
\eea

At Born level, $\cl{O}(\as^0)$, we find
\beq
\hat{\s}_{q\bar{q}}^{(0)} = \frac{1}{2x\xpr s}\frac{1}{N_c}\left[2\pi\delta(x\xpr s-Q^2)\right] \left[4\pi\al_{q\bar{q}} Q^2\right] = \frac{4\pi^2\al_{q\bar{q}}}{N_c s} \delta(x\xpr-\tau)\;,
\eeq
where $4\pi\al_{q\bar{q}}$ is the coupling of the quark $q$ and antiquark $\bar{q}$ to the vector boson, which in the case of $W$ production includes the relevant CKM matrix elements.

The $\cl{O}(\as)$ correction comprises contributions from real and virtual gluon emission. Denoting the 4-momenta of the incoming quark and antiquark $p$ and $p^\prime$, that of the gluon $k$ and the vector boson $q$, we may write in the centre-of-momentum frame of the incoming quark and antiqark,
\beq
p=\frac{\sqrt{x\xpr s}}{2}(1,0,0,1)\ind p^\prime=\frac{\sqrt{x\xpr s}}{2}(1,0,0,-1)
\eeq
and
\beq
k=(k^0,k_\perp\cos\phi,k_\perp\sin\phi,k^3)\;.
\eeq

To evaluate the real contribution, we note that the squared matrix element is given by
\beq
e^{i\bv\cdot\kp}\vert M\vert^2 = e^{i\bv\cdot\kp}\frac{1}{N_c}\left[4\pi\al_{q\bar{q}}\right]\left(8\pi\as C_F\right)\frac{(p\cdot q)^2+(p^\prime\cdot q)^2}{(p\cdot k)(p^\prime\cdot k)}\;,
\eeq
which, in terms of the variables
\beq
z=\frac{\tau}{x\xpr}\ind \D=\frac{k_\perp^2}{Q^2}\;,
\eeq
may be written
\beq
e^{i\bv\cdot\kp}\vert M\vert^2 = e^{ibQ\sqrt{\D}\cos\phi}\frac{4\pi}{z\D}\frac{8\pi^2\al_{q\bar{q}}}{N_c}\frac{\as C_F}{2\pi}\left[2+2z^2-4z\D\right]\;.
\eeq
The Lorentz-invariant integration measure and phase space for the squared matrix elements is
\beq
\frac{\Theta(1-z)}{4\pi}\int_0^{(1-z)^2/(4z)}\frac{z\,d\D}{\sqrt{(1-z)^2-4z\D}}\int_0^{2\pi}\frac{d\phi}{2\pi}\;,
\eeq
and thus we obtain the $\cl{O}(\as)$ real contribution as
\beq
\hat{\s}_{q\bar{q}}^\sml{real} = \frac{4\pi^2\al_{q\bar{q}}}{N_c s}\frac{\Theta(1-z)}{x\xpr}\frac{\as C_F}{2\pi}\int_0^{(1-z)^2/(4z)}\frac{d\D}{\D}\,J_0(bQ\sqrt{\D})\frac{2+2z^2-4z\D}{\sqrt{(1-z)^2-4z\D}}\;.
\eeq
This is divergent at small $\D$ when the gluon becomes soft or collinear with one of the incoming quarks; we will therefore find it convenient to separate this logarithmic divergence from the finite remainder.

In order to simplify the calculation we take the Mellin transform
\beq
\Sgt(N,p_\perp^2) \equiv \int_0^1 d\tau\,\tau^{N-1}\Sigma(\tau,p_\perp^2)\;.
\eeq
This yields
\beq
\Sgt(N,p_\perp^2) = \frac{1}{2}\int_0^\infty db\,bJ_0(bp_\perp)\sum_{q\bar{q}}\frac{4\pi^2\al_{q\bar{q}}}{N_c s}\left[q_{N}^{}\bar{q}_{N}^\prime+\bar{q}_{N}^{}q_{N}^\prime\right]\left\{1+C_1(N,b)-R(N,b)\right\}\;,
\eeq
where in the final term on the right hand side, the 1 represents the Born level contribution, and the quantity $C_1-R$ the finite and divergent parts of the $\cl{O}(\as)$ correction. The moments of the parton distributions are
\beq
q_N = \int_0^1 dx \, x^{N-1} q(x) \;.
\eeq
Thus we obtain for the real gluon emission:
\beq
\label{Creal}
C_1^\sml{real}-R^\sml{real} = \frac{\as C_F}{2\pi}\int_0^\infty\frac{d\D}{\D}J_0(bQ\sqrt{\D})\int_0^{1+2\D-2\sqrt{\D(1+\D)}}dz\,z^{N-1}\frac{2+2z^2-4z\D}{\sqrt{(1-z)^2-4z\D}}\;.
\eeq

The $z$ integral may be evaluated as a series expansion in $\D$ (see the appendix for details) to give the collinear and soft divergences:
\beq
\label{apc3}
R^\sml{real}(N,b) = \frac{\as C_F}{2\pi}\int_0^1\frac{d\D}{\D}J_0(bQ\sqrt{\D})\left\{2\log\D+4S_1(N)+\frac{2}{N}+\frac{2}{N+1}\right\}\;.
\eeq

Returning now to the virtual contribution, by expressing the loop integrals as a single integral over $k_\perp$, we obtain
\beq
\hat{\s}_{q\bar{q}}^\sml{virt} = \frac{4\pi^2\al_{q\bar{q}}}{N_c s} \delta(xx^\prime-\tau)\frac{\as C_F}{2\pi}\int_0^\infty\frac{d\D}{\D}\nml{Re}\left\{3+\frac{2-6\D}{\sqrt{1-4\D}}\log\frac{\sqrt{1-4\D}-1}{\sqrt{1-4\D}+1}\right\}\;.
\eeq
This also diverges at the lower limit, and thus for the purposes of a full $\cl{O}(\as)$ calculation a regulator is required, such as dimensional regularisation or, as performed in section \ref{powcor}, a small gluon mass. But this expression is sufficient for obtaining the logarithmic divergences. The virtual contribution is separated into finite and divergent pieces according to:
\beq
\hat{\s}_{q\bar{q}}^\sml{virt} = \frac{4\pi^2\al_{q\bar{q}}}{N_c s} \delta(xx^\prime-\tau)\left[C_1^\sml{virt}-R^\sml{virt}\right]\;,
\eeq
where the collinear divergence $R$ is found by expanding the integrand in $\D$:
\beq
R^\sml{virt} = \frac{\as C_F}{2\pi}\int_0^1\frac{d\D}{\D}\Bigl\{-2\log\D-3\Bigr\}\;.
\eeq

Thus the sum of real and virtual divergences becomes:
\beq
R = \frac{\as C_F}{2\pi}\int_0^{Q^2}\frac{dk_\perp^2}{k_\perp^2}\left\{-2\left[1-J_0(bk_\perp)\right]\log\frac{k_\perp^2}{Q^2}-3+2J_0(bk_\perp)\left(2S_1(N)+\frac{1}{N}+\frac{1}{N+1}\right)\right\}\;.
\eeq
There is a certain freedom in the choice of $R$, notably in the choice of the upper limit $k_\perp^2=Q^2$. But this is compensated by the finite contribution $C_1$, which is defined such that $C_1-R$ gives the full $\cl{O}(\as)$ virtual contribution.

The radiator $R$ contains the logarithmic divergences associated with single-gluon emission. It is a non-trivial function of $N$, but can be simplified by a wise choice of factorisation scale. We begin by splitting the radiator into three pieces
\beq
R = R_1 + R_2 + R_3 \;,
\eeq
where
\bea
R_1 &=& \frac{\as C_F}{2\pi}\int_0^{Q^2}\frac{dk_\perp^2}{k_\perp^2}\left[1-J_0(bk_\perp)\right]\left(-2\log\frac{k_\perp^2}{Q^2}-3\right)\\
R_2 &=& \frac{\as C_F}{2\pi}\int_0^{Q^2}\frac{dk_\perp^2}{k_\perp^2}\left[1-J_0(bk_\perp)\right]\left(3-4S_1(N)-\frac{2}{N}-\frac{2}{N+1}\right)\\
R_3 &=& \frac{\as C_F}{2\pi}\int_0^{Q^2}\frac{dk_\perp^2}{k_\perp^2}\left(4S_1(N)+\frac{2}{N}+\frac{2}{N+1}-3\right)\;.
\eea

The first two pieces may be simplified by making the {\it de facto} approximation of the Bessel function:
\beq
\label{besappr}
\left[1-J_0(bk_\perp)\right] \approx \Theta(bk_\perp-b_0)\;,
\eeq
where $b_0=2e^{-\gamma_E}$ and $\gamma_E$ is Euler's constant, as described more fully in the appendix. Thus
\bea
R_1 &=& \frac{\as C_F}{2\pi}\int_{(b_0/b)^2}^{Q^2}\frac{dk_\perp^2}{k_\perp^2}\left(-2\log\frac{k_\perp^2}{Q^2}-3\right)\\
R_2 &=& \frac{\as C_F}{2\pi}\int_{(b_0/b)^2}^{Q^2}\frac{dk_\perp^2}{k_\perp^2}\left(3-4S_1(N)-\frac{2}{N}-\frac{2}{N+1}\right)
\eea

The contribution $R_3$ is formally divergent. Using dimensional regularisation, with $4+2\eta$ dimensions, it becomes
\beq
\label{R3dimreg}
R_3 = \frac{\as C_F}{2\pi}\int_0^{Q^2}\frac{dk_\perp^2}{k_\perp^2}\frac{(4\pi\mu^2/k_\perp^2)^{-\eta}}{\Gamma(1+\eta)}\left(4S_1(N)+\frac{2}{N}+\frac{2}{N+1}-3+\cl{O}(\eta)\right)
\eeq
where $\mu$ is the factorisation scale which appears in the quark distributions. On application of \MSbar, we obtain
\beq
R_3 = \frac{\as C_F}{2\pi}\int_{\mu^2}^{Q^2}\frac{dk_\perp^2}{k_\perp^2}\left(4S_1(N)+\frac{2}{N}+\frac{2}{N+1}-3\right)\;,
\eeq
where the finite contribution arising from the $\cl{O}(\eta)$ term in equation \rf{R3dimreg} has been absorbed into $C_1$. Let us therefore choose the factorisation scale to be $\mu^2=(b_0/b)^2$, in order that the terms $R_2$ and $R_3$ cancel. Alternatively, since these contributions are proportional to the non-singlet anomalous dimension, starting from any factorisation scale $\mu^2$ the terms $R_2$ and $R_3$ may be absorbed into the parton densities by means of a scale change to $(b_0/b)^2$.

Thus the single-gluon contribution may be written as
\bea
\Sgt(N,p_\perp^2) &=& \frac{1}{2}\int_0^\infty db\,bJ_0(bp_\perp)\sum_{q\bar{q}}\frac{4\pi^2\al_{q\bar{q}}}{N_c s}\left[q_{N}^{}((b_0/b)^2)\bar{q}_{N}^\prime((b_0/b)^2)+(\leftrightarrow)\right]\nln
& & \times\left\{1+C_1(N,b)-R(b)\right\}\;,
\eea
where the collinear and soft logarithms are represented by the radiator
\beq
R = \frac{\as C_F}{2\pi}\int_{(b_0/b)^2}^{Q^2}\frac{dk_\perp^2}{k_\perp^2}\left(-2\log\frac{k_\perp^2}{Q^2}-3\right)\;,
\eeq
and the remaining $\cl{O}(\as)$ contribution is to be found in $C_1$.

\subsect{Multiple gluon emission}
Consider now the case in which many gluons are emitted before the vector boson is produced. Each splitting contributes to the total transverse momentum of the event: we are interested only in the contributions from soft and collinear logarithms, which may be calculated using the collinear approximation.

Let the 4-momenta of the incident hadrons be $P$ and $P^\prime$. In the centre of momentum frame we may write
\beq
P = \frac{\sqrt{s}}{2}(1,0,0,1) \hspace{0.7in}
P^\prime = \frac{\sqrt{s}}{2}(1,0,0,-1) \;,
\eeq
where $s$ is the total energy of the collision.

Suppose the (anti)quark in $P$ has momentum fraction $x$; and let it radiate $n$ gluons with momenta $k_i$, $(i=1,\cdots,n)$. We may write the Sudakov decomposition
\beq
k_i = \al_i P + \be_i P^\prime + \kp_i \;,
\eeq
in terms of which the integration measure is
\beq
\int\frac{d^4k_i}{(2\pi)^4}\,2\pi\delta^+(k_i^2) = \int\frac{d\al_i}{2\al_i}\frac{d^2\kp_i}{(2\pi)^3}\;.
\eeq

Let $z_i$ be the fraction of remaining longitudinal momentum retained by the (anti)quark at the $i$th splitting, i.e.
\bea
\al_i &=& (1-z_i)\left(x-\sum_{j=1}^{i-1}\al_j\right) \\
z_i &=& \frac{x-\sum_{j=1}^{i}\al_j}{x-\sum_{j=1}^{i-1}\al_j} \;.
\eea
Then the collinear-divergent pieces of the squared matrix element are given by the usual splitting function:
\beq
\vert\cl{M}\vert_{i-1}^2 = \frac{\as C_F}{2\pi}\int\frac{dk_{\perp i}^2}{k_{\perp i}^2}\frac{d\phi_i}{2\pi}\frac{dz_i}{z_i}\frac{1+z_i^2}{1-z_i}\vert\cl{M}\vert_i^2 e^{ibk_{\perp i}\cos\phi}\;,
\eeq
where $\phi_i$ is the azimuthal angle of the $i$th gluon. Since we are considering collinear emission, the phase space is subject to the condition $\al_i > \be_i$, which translates in our approximation to
\beq
z < 1 - \frac{k_\perp}{Q}\;.
\eeq
This boundary is not the boundary of physical phase space; rather it is imposed as part of the collinear approximation. It is necessary therefore to check that any dependence in our calculation on this limit is a genuine physical effect and not an artefact of the approximation.

We may treat radiation from the other incoming (anti)quark in the same way, but with $\al_i$ and $\be_i$ interchanged. The real part of the partonic cross-section appeaing in \rf{eqSgm} is then:
\bea
\label{sigma}
\lefteqn{\hat{\sigma}_{q\bar{q}}^\sml{real} = \frac{1}{2x\xpr s}\frac{1}{N_c}\left[4\pi\al_{q\bar{q}}\tau\right]\sum_{n=0}^{\infty}\frac{1}{n!}\prod_{i=1}^{n}\left\{\frac{\as C_F}{2\pi}\int\frac{dk_{\perp i}^2}{k_{\perp i}^2}J_0(bk_{\perp i})\int\frac{dz_i}{z_i}\frac{1+z_i^2}{1-z_i}\right\}\times}\\
& & \sum_{n^\prime=0}^{\infty}\frac{1}{n^\prime!}\prod_{i^\prime=1}^{n^\prime}\left\{\frac{\as C_F}{2\pi}\int\frac{dk_{\perp i^\prime}^{\prime 2}}{k_{\perp i^\prime}^{\prime 2}}J_0(bk_{\perp i^\prime}^\prime)\int\frac{dz_{i^\prime}^\prime}{z_{i^\prime}^\prime}\frac{1+z_{i^\prime}^{\prime 2}}{1-z_{i^\prime}^\prime}\right\}\,2\pi\delta(x\xpr z_1\cdots z_n z_1^\prime\cdots z_{n^\prime}^\prime-\tau)\;.\nonumber
\eea

Taking the Mellin transform leads to:
\beq
\Sgt(N,p_\perp^2) = \frac{1}{2}\int_0^\infty db\,bJ_0(bp_\perp)\sum_{q\bar{q}}\frac{4\pi^2\al_{q\bar{q}}}{N_c s}\left[q_{N}^{}\bar{q}_{N}^\prime+\bar{q}_{N}^{}q_{N}^\prime\right]e^{-R(N,b)}\;,
\eeq
where the real emission radiator is
\beq
\label{Rreal}
R_\sml{real}(N,b) = -\frac{\as C_F}{\pi}\int_0^{Q^2}\frac{dk_\perp^2}{k_\perp^2}J_0(bk_\perp)\int_0^{1-k_\perp/Q}dz\,z^{N-1}\frac{1+z^2}{1-z}\;.
\eeq

We then take into account virtual gluon radiation using the universal virtual contribution to the radiator \cite{eec}
\beq
\label{Rvirt}
R_\sml{virtual}(b) = \frac{\as C_F}{\pi}\int_0^{Q^2}\frac{dk_\perp^2}{k_\perp^2}\int_0^{1-k_\perp/Q}dz\frac{1+z^2}{1-z}\;,
\eeq
so that the full radiator is
\beq
\label{rad}
R(N,b) = \frac{\as C_F}{\pi}\int_0^{Q^2}\frac{dk_\perp^2}{k_\perp^2}\int_0^{1-k_\perp/Q}dz\frac{1+z^2}{1-z}\left[1-z^{N-1}J_0(bk_\perp)\right]\;.
\eeq
Performing the $z$ integral at small $k_\perp^2$ yields:
\beq
R(N,b) = \frac{\as C_F}{2\pi}\int_0^{Q^2}\frac{dk_\perp^2}{k_\perp^2}G(k_\perp^2/Q^2, bk_\perp, N)\;.
\eeq
where
\bea
\label{Gint}
G(k_\perp^2/Q^2,bk_\perp,N) &=& -2\left[1-J_0(bk_\perp)\right]\log\frac{k_\perp^2}{Q^2}-3+2J_0(bk_\perp)\left(2S_1(N)+\frac{1}{N}+\frac{1}{N+1}\right)\nln
& & \hspace{2.0cm}+4\left[1-NJ_0(bk_\perp)\right]\frac{k_\perp}{Q}+\cl{O}\left(\frac{k_\perp^2}{Q^2}\right)\;.
\eea

It should be noted that:
\begin{itemize}
\item This result is strictly valid provided $N$ is not too large, which here means approximately $N<Q/k_\perp$, since as $N\to\infty$ the contribution from real emission in \rf{rad} vanishes. But we are not interested in such large values of $N$, since they correspond to values of $\tau$ close to $1$, while for vector boson production at the LHC or Tevatron the order of magnitude is $\tau\sim0.01$.
\item The terms of order $k_\perp^2/Q^2$ do not generate logarithmic divergences: they are thus included in the finite contribution to the cross-section and therefore do not feature in the radiator. The term of order $k_\perp/Q$ is a spurious artefact of the collinear approximation, that is not present in the single-gluon result. The collinear approximation is thus valid only for logarithm resummation, and may not be used for determinations of power corrections \cite{bebr}.
\end{itemize}

The decomposition of the radiator proceeds just as in the single-gluon case. The factorisation scale is naturally chosen to be $\mu^2=(b_0/b)^2$ and we obtain the exponentiated form:
\beq
\Sgt(N,p_\perp^2) = \frac{1}{2}\int_0^\infty db\,bJ_0(bp_\perp)\sum_{q\bar{q}}\frac{4\pi^2\al_{q\bar{q}}}{N_c s}\left[q_{N}^{}((b_0/b)^2)\bar{q}_{N}^\prime((b_0/b)^2)+(\leftrightarrow)\right]C(\as)e^{-R(b)}\;,
\eeq
where the leading collinear and soft logarithms are represented by the radiator
\beq
R = \frac{\as C_F}{2\pi}\int_{(b_0/b)^2}^{Q^2}\frac{dk_\perp^2}{k_\perp^2}\left(-2\log\frac{k_\perp^2}{Q^2}-3\right)\;,
\eeq
and the remaining contribution to fixed order in perturbation theory is given by the factor
\beq
C(\as)=1+C_1(N,b)+\cl{O}(\as^2)\;.
\eeq

Next-to-leading logarithms have been calculated \cite{nll,elve,css,fnr}, and may be included in the radiator by writing
\beq
R = \frac{C_F}{2\pi}\int_{(b_0/b)^2}^{Q^2}\frac{dk_\perp^2}{k_\perp^2}\as^\sml{PT}(k_\perp^2)\left(-2\log\frac{k_\perp^2}{Q^2}-3\right)\;,
\eeq
where the perturbative coupling is defined in the Bremsstrahlung scheme \cite{cmw}, and is, to next-to-leading order,
\beq
\label{asPT}
\as^{\sml{PT}}(k_\perp^2) = \as^\smMSb(Q^2) + \left[\as^\smMSb(Q^2)\right]^2\left(\frac{\be_0}{4\pi}\log\frac{Q^2}{k_\perp^2}+\frac{K}{4\pi}\right)\;,
\eeq
where
\beq
K = \frac{(67-3\pi^2)C_A-20T_R n_f}{9}\;.
\eeq

Having now recalled the perturbative contribution to the transverse momentum distribution, we are in a position to calculate non-perturbative effects from the modification to the coupling at low scales.

\sect{Non-perturbative contribution}
\label{powcor}
The dispersive approach to power corrections \cite{dmw} has been applied to a variety of QCD observables. The basic postulate is that we may define a universal running strong coupling $\as(k^2)$ that is well-behaved in the infra-red as well as in the ultra-violet, having no singularities in the complex plane apart from a branch cut along the negative real axis. We may then write a dispersion relation
\beq
\as(k^2) = -\int_0^\infty\frac{d\mu^2}{k^2+\mu^2}\rho_s(\mu^2)\;,
\eeq
where $\rho_s(\mu^2)$ is the discontinuity of $\as$ across its branch cut:
\beq
\rho_s(\mu^2) = \frac{1}{2\pi i}\left\{\as(e^{i\pi}\mu^2)-\as(e^{-i\pi}\mu^2)\right\}\;.
\eeq

Non-perturbative contributions to the transverse momentum distribution arise from the modification to the running coupling at low scales:
\beq
\as(k^2) = \as^\sml{PT}(k^2)+\de\as(k^2)\;,
\eeq
where the contribution $\as^\sml{PT}$, given in \rf{asPT}, is that already accounted for in the perturbative part of the calculation. In the dispersive approach the corresponding modification to the spectral function, $\delta\rho_s(k^2)$, gives the correction to the coefficient function
\beq
\delta C(N,b) = \frac{C_F}{2\pi}\int_0^\infty\frac{dk^2}{k^2}\de\rho_s(k^2)\left[\cl{C}(k^2/Q^2)-\cl{C}(0)\right]
\eeq
where $(\as C_F/2\pi)\cl{C}(k^2/Q^2)$ is just the single-gluon contribution to $C(N,b)$ calculated as though gluon has mass $k^2$. As customary we define $\e=k^2/Q^2$.

The coefficient function has contributions from both real and virtual gluons. Consider first the real contribution. A calculation of the massive single-gluon matrix element and phase space gives
\beq
\cl{C}^\sml{real} = \int_0^\infty\frac{d\D}{\D}J_0(bQ\sqrt{\D})\int_0^{Z(\D,\e)}dz\,z^{N-1}\frac{2\D(\D+z^2\D-2z\D^2+2\e z^2+2\e^2 z^2+\e^2 z^2\D)}{(\D+\e z)^2\sqrt{(1-z+\e z)^2-4z(\e+\D)}}
\eeq
where the upper limit on the $z$ integral is
\beq
Z(\D,\e) = \frac{(1+2\D+\e)-2\sqrt{(\e+\D)(1+\D)}}{(1-\e)^2}
\eeq
Note how this reduces to perturbative case \rf{Creal} when we put $\e=0$.

In order to proceed with this integral we expand the Bessel function. We require only those terms that are non-analytic in $\e$ as $\e\to0$, and these are generated at the lower limit of the $\D$ integral. So the expansion is legitimate for our purposes and we obtain
\bea
\cl{C}^\sml{real}(\e) &=& \int_0^{1/(1+\sqrt{\e})^2}dz\,z^{N-1}\int_0^{(1-z+\e z)^2/(4z)-\e}d\D\left(1-\frac{b^2Q^2\D}{4}+\cdots\right)\times\nln
& & \ind\frac{2(\D+z^2\D-2z\D^2+2\e z^2+2\e^2 z^2+\e^2 z^2\D)}{(\D+\e z)^2\sqrt{(1-z+\e z)^2-4z(\e+\D)}}\;.
\eea

In order to evaluate these integrals we follow the method of \cite{dmw}. The integral over $\D$ is relatively straightforward, giving
\bea
\cl{C}^\sml{real}(\e) &=& \int_0^{1/(1+\sqrt{\e})^2}dz\,z^{N-1}\Biggl[\frac{P(z,\e)}{1-z-\e z}\tanh^{-1}\frac{\sqrt{(1-z-\e z)^2-4\e z^2}}{1-z-\e z}\nln
& & \ind-Q(z,\e)\sqrt{(1-z-\e z)^2-4\e z^2}\Biggr]\;,
\eea
where the functions $P$ and $Q$ are given by
\bea
P(z,\e) &=& 4+4(1+\e)^2 z^2 + 2\e b^2 Q^2 z (1-z+z^2-\e z+2\e z^2+\e^2 z^2)\\
Q(z,\e) &=& 4+\frac{b^2 Q^2}{6z}(1+z+z^2+\e z+10\e z^2+\e^2 z^2)\;.
\eea
We then make the substitution
\beq
\eta = \tanh^{-1}\frac{\sqrt{(1-z-\e z)^2-4\e z^2}}{1-z-\e z}\ind
z=\frac{1}{1+\e+2\sqrt{\e}\cosh\eta}\;,
\eeq
which leads to the expression for the characteristic function
\bea
\lefteqn{\cl{C}^\sml{real}(\e) = 4I_N + 2\e b^2 Q^2 I_{N+1} + 4(1+\e)^2 I_{N+2} - 2\e(1+\e)b^2 Q^2 I_{N+2}+2\e(1+\e)^2 b^2 Q^2 I_{N+3}}\nln
& & -{\textstyle\frac{2}{3}}\e b^2 Q^2 J_{N+1}-16\e J_{N+2}-{\textstyle\frac{2}{3}}\e(1+\e)b^2 Q^2 J_{N+2}-{\textstyle\frac{2}{3}}\e(1+10\e+\e^2)b^2 Q^2 J_{N+3}\;,\hspace{1.0cm}
\eea
in terms of the integrals
\bea
\label{IJints}
I_N(\e) &=& \int_0^\infty d\eta\frac{\eta\tanh\eta}{(1+\e+2\sqrt{\e}\cosh\eta)^N}\nln
J_N(\e) &=& \int_0^\infty d\eta\frac{\sinh^2\eta}{(1+\e+2\sqrt{\e}\cosh\eta)^N}\;.
\eea
Using the values of these integrals given in the appendix, we obtain the contribution from real gluon emission as:
\bea
\cl{C}^\sml{real}(\e) &=& \left\{\pi^2+(2S_1(N)+\log\e)^2-4S_2(N)+2\left(\frac{1}{N}+\frac{1}{N+1}\right)(2S_1(N)+\log\e)\right\}\nln
& & +\frac{b^2Q^2}{6}\left(\frac{1}{N+2}-\frac{1}{N-1}\right)\nln
& & +\e\left\{2(2S_1(N)+\log\e)\left(N^2+3N-3+\frac{1}{N+1}\right)-N\pi^2\right.\nln
& & \ind\left.-N(2S_1(N)+\log\e)^2+4NS_2(N)-6N^2-6N+16-\frac{8}{N+1}\right\}\nln
& & + \frac{\e b^2Q^2}{4}\left\{\pi^2+(2S_1(N)+\log\e)^2-4S_2(N)+2-\frac{6}{N+2}\right.\nln
& & \ind\left.+2\left(\frac{2}{N}+\frac{2}{N+2}-1\right)(2S_1(N)+\log\e)\right\}+\cdots\;.
\eea

Now we turn to the virtual contribution. Since the virtual gluon does not contribute to the transverse momentum of the vector boson, we obtain the same as that calculated in \cite{dmw} for the total Drell-Yan cross-section, which is
\bea
\cl{C}^\sml{virt} &=& (1+\e)^2\left\{2\nml{Li}_2(-\e)+2\log\e\log(1+\e)+\frac{\pi^2}{3}-\log^2\e\right\}-\frac{7}{2}-2\e-(3+2\e)\log\e\nln
&=& \left(-\log^2\e-3\log\e+\frac{\pi^2}{3}-\frac{7}{2}\right)+\e\left(-2\log^2\e+\frac{2\pi^2}{3}-4\right)+\cl{O}(\e^2)\;.
\eea

The power correction is most conveniently expressed in terms of moments of the modification to the coupling, $\de\as(k^2)$:
\beq
\de C(N,b)=\frac{C_F}{2\pi}\int_0^\infty\frac{dk^2}{k^2}\de\as(k^2)\cl{D}(k^2/Q^2)\;,
\eeq
where $\cl{D}(\e)$ is the discontinuity of $\cl{C}(\e)$ across its branch cut:
\beq
\cl{D}(\e) = -\frac{1}{2\pi i}\left[\cl{C}(\e e^{i\pi})-\cl{C}(\e e^{-i\pi})\right].
\eeq
The leading term in the characteristic function is the $\log\e$ divergence proportional to the anomalous dimension: this is subtracted off and absorbed into the parton density functions. From the finite contribution we obtain
\bea
\cl{D}(\e) &=& -2\e\left\{(N+2)\log\e+2NS_1(N)-N^2-3N+3-\frac{1}{N+1}\right\}\nln
& & \ind+\frac{\e b^2Q^2}{2}\left\{\log\e+2S_1(N)+\frac{2}{N}+\frac{2}{N+2}-1\right\}+\cdots\;.
\eea
The $b$-independent term on the first line gives the $1/Q^2$ corrections to the total cross-section calculated in \cite{dmw}. Of particular relevance to the transverse momentum distribution is the $b$-dependent term on second line, which gives the correction
\beq
\label{eqcorr}
\delta C(N,b) = -\frac{b^2}{2}\left\{\cl{A}_2\left(\log Q^2-2S_1(N)-\frac{2}{N}-\frac{2}{N+2}+1\right)-\cl{A}_2^\prime\right\}\;.
\eeq
The non-perturbative parameters $\cl{A}_2$ and $\cl{A}_2^\prime$ are defined in terms of $\delta\as$ by:
\bea
\cl{A}_2 &=& \frac{C_F}{2\pi}\int_0^\infty\frac{dk_\perp^2}{k_\perp^2}\,k_\perp^2\delta\as(k_\perp^2) \\
\cl{A}_2^\prime &=& \frac{C_F}{2\pi}\int_0^\infty\frac{dk_\perp^2}{k_\perp^2}\,k_\perp^2\log k_\perp^2 \delta\as(k_\perp^2)\;.
\eea
The extent of our knowledge of $\cl{A}_2$ and $\cl{A}_2^\prime$ is \cite{eec,dissf} that $\cl{A}_2 \approx 0.2$ GeV$^2$ and $\cl{A}_2^\prime$ is at present unconstrained but appears consistent with zero.

\sect{Results and discussion}
\label{result}
In the previous section we derived the non-perturbative 1-gluon contribution to the transverse momentum distribution in impact parameter space, which by performing the inverse Mellin transform may be written:
\beq
\label{neint}
\Sg(\tau,p_\perp^2) = \frac{1}{2}\int_0^\infty db\,bJ_0(bp_\perp)e^{-R(b)}K(\tau,b)\left[1-b^2 W(\tau,b)\right]
\eeq
where the perturbative contribution is
\beq
K(\tau,b) = \sum_{q\bar{q}}\frac{4\pi^2\al_{q\bar{q}}}{N_c s}\int_0^1 dxdx^\prime \left[q(x,(b_0/b)^2)\bar{q}^\prime(x^\prime,(b_0/b)^2)+(\leftrightarrow)\right]\de(xx^\prime-\tau)\;,
\eeq
and the non-perturbative component is
\beq
W(\tau,b) = \frac{1}{2}\left\{\cl{A}_2\log Q^2+\cl{A}_2-\cl{A}_2^\prime\right\}-\frac{\cl{A}_2}{K(\tau,b)}\int_\tau^1\frac{dz}{z}K(\tau/z,b)\left(1+z^2-\frac{1}{(1-z)_+}\right)\;.
\eeq

\begin{figure}[ht]
\begin{center}
\epsfig{file=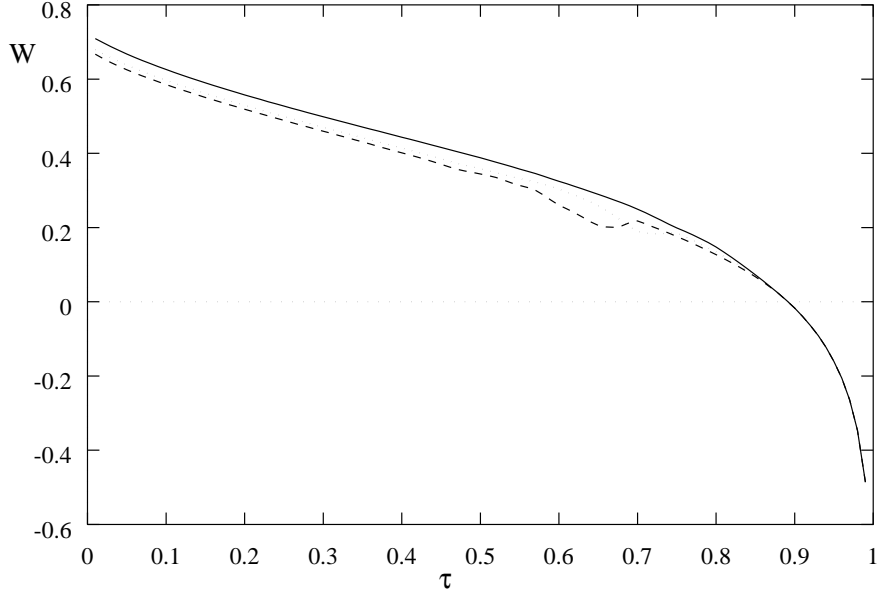}
\caption{\label{graph}Graph of $W(\tau,b)$ against $\tau$ for $b=1$ GeV$^{-1}$ (solid line), $b=0.1$ GeV$^{-1}$ (dotted) and $b=0.01$ GeV$^{-1}$ (dashed), evaluated at the Z mass with $\cl{A}_2=0.2$ GeV$^2$ and $\cl{A}_2^\prime=0$.}
\end{center}
\end{figure}

Figure \ref{graph} shows a plot of $W(\tau,b)$ against $\tau$ at fixed values of $b$, for on-shell Z$^0$ production with $\cl{A}_2=0.2$ GeV$^2$, $\cl{A}_2^\prime=0$, evaluated using the MRST (central gluon) distributions \cite{MRST}; the slight dependence of $W(\tau,b)$ on $b$ is logarithmic, coming only from the scale of the parton distributions. It is also clear from the figure that the result given here is not valid near $\tau=1$: it becomes negative and diverges logarithmically. This is because our expression for the Mellin transform was valid for $N$ not too large (equation \rf{Gint} in the perturbative case, and analogously though not so obviously in the non-perturbative calculation), i.e. $N<Q/k_\perp$, while as $\tau\to 1$ the large moments become important. The approximation is valid away from this region, and is certainly legitimate for LHC or Tevatron data.

We thus predict a non-perturbative correction that behaves quadratically in $b$ near $b=0$, in agreement with existing models based on a Gaussian suppression factor \cite{css}. We can make our result Gaussian also by exponentiating:
\beq
1-b^2 W(\tau,b) \approx e^{-b^2 W(\tau,b)}\;.
\eeq
Whether or not the power correction does exponentiate is an open question --- intuitively it seems reasonable, since we expect contributions from all soft gluons, not just the gluon nearest the hard vertex; yet it is not proven and should be treated with care, since exponentiation normally applies to collinear and soft logarithms, while the power corrections come from terms that converge in the infra-red. Nevertheless there are several good reasons for at least studying the exponentiated correction. Firstly, the integral \rf{neint} diverges at large $b$ due to the non-perturbative contribution growing as $b^2$, while the exponentiated expression is convergent. At small $b$, where the non-perturbative correction is small, whether or not we choose to exponentiate makes minimal difference.

Secondly, we can consider the difference between the exponentiated and un-ex\-po\-nen\-ti\-ated expressions to give a measure of the theoretical uncertainty in our calculation, since we expect higher orders to contribute at approximately this level. Thus at small $b$ the error is very small, while at large $b$ it becomes huge. This has the same pattern as the general impact parameter formalism, which suffers from considerable uncertainties at large $b$ due to the parton distributions becoming unknown.

Finally, there have been some fairly reasonable phenomenological fits of Gaussian models to data, such as performed in \cite{lbly}. A two-parameter fit yielded the non-perturbative contribution to the exponent
\beq
\label{tpf}
-b^2\left[0.34\nml{ GeV}^2\log\left(\frac{Q}{1\nml{ GeV}}\right)-0.16\nml{ GeV}^2\right]\;,
\eeq
while a three-parameter fit, with an additional linear term, may be written
\beq
\label{thpf}
-b^2\left[0.48\nml{ GeV}^2\log\left(\frac{Q}{1\nml{ GeV}}\right)-0.41\nml{ GeV}^2\right]+b\left[0.09\nml{ GeV}\log(100x\xpr)\right]\;.
\eeq
These may be compared with the prediction from the dispersive approach, which is
\beq
-b^2 W = -b^2\left[0.2\nml{ GeV}^2\log\left(\frac{Q}{1\nml{ GeV}}\right)-0.2\nml{ GeV}^2\right]\;\;\;\nml{ at }\;\;\;\tau=0.01\;,
\eeq
where we have used $\cl{A}_2=0.2$ GeV$^2$ and $\cl{A}_2^\prime=0$. It is seen that this accounts for a substantial portion of the total non-perturbative function fitted to the data; however the $\tau$-dependence of the fitted forms \rf{tpf} and \rf{thpf} differs so greatly from that predicted here that it is hard to give more than this qualitative statement. A full fit to data is beyond the scope of this current article, but such a fit would be a further test of the dispersive model of power corrections, would enable one to extract a value for the unknown non-perturbative parameter $\cl{A}_2^\prime$, and would test the assumption of universality which predicts $\cl{A}_2=0.2$ GeV$^2$ from DIS structure functions \cite{dissf}.

An alternative approach to the non-perturbative contribution \rf{eqcorr} would be to interpret part of it as a shift in the scale at which the parton distributions are evaluated. We may write the contribution as
\beq
\label{eqadc}
\de C(N,b) = -\frac{b^2\cl{A}_2}{2C_F}\g_{qq}(N)-\frac{b^2}{2}\left\{\cl{A}_2\left(\log Q^2-\frac{1}{N}+\frac{1}{N+1}-\frac{2}{N+2}-\frac{1}{2}\right)-\cl{A}_2^\prime\right\}\;,
\eeq
where the non-singlet anomalous dimension is
\beq
\g_{qq}(N)=C_F\int_0^1 dz\,z^{N-1}\left(\frac{1+z^2}{1-z}\right)_+=C_F\left(\frac{3}{2}-2S_1(N)-\frac{1}{N}-\frac{1}{N+1}\right)\;.
\eeq
The first term on the right hand side of \rf{eqadc} then shifts the factorisation scale from $(b_0/b)^2$ to $(b_0/b)^2+\de\mu^2$, where
\beq
-\frac{b^2\cl{A}_2}{2C_F}\g_{qq}(N) = 2\int_{(b_0/b)^2}^{(b_0/b)^2+\de\mu^2}\frac{dk_\perp^2}{k_\perp^2}\frac{\as(k_\perp^2)}{2\pi}\g_{qq}(N)\;.
\eeq
If as a first approximation we replace the integrand by its value at the lower limit, we obtain the shift
\beq
\de\mu^2 = -\frac{b_0^2\cl{A}_2}{4}\left[\frac{\as((b_0/b)^2)C_F}{2\pi}\right]^{-1}\approx -3\mbox{ GeV}^2\;,
\eeq
where in obtaining the approximate numerical value, the running of $\as$ is neglected.

So a negative correction to the factorisation scale is predicted, in addition to which there is the remainder of the non-perturbative contribution calculated above. Thus instead of \rf{neint}, we may write
\beq
\Sg(\tau,p_\perp^2) = \frac{1}{2}\int_0^\infty db\,bJ_0(bp_\perp)e^{-R(b)}\hat{K}(\tau,b)\left[1-b^2\hat{W}(\tau,b)\right]
\eeq
where the perturbative contribution is now
\beq
\hat{K}(\tau,b) = \sum_{q\bar{q}}\frac{4\pi^2\al_{q\bar{q}}}{N_c s}\int_0^1 dxdx^\prime \left[q(x,(b_0/b)^2+\de\mu^2)\bar{q}^\prime(x^\prime,(b_0/b)^2+\de\mu^2)+(\leftrightarrow)\right]\de(xx^\prime-\tau)\;,
\eeq
and the non-perturbative component is
\beq
\hat{W}(\tau,b) = \frac{1}{4}\left\{2\cl{A}_2\log Q^2-\cl{A}_2-2\cl{A}_2^\prime\right\}-\frac{\cl{A}_2}{2\hat{K}(\tau,b)}\int_\tau^1\frac{dz}{z}\hat{K}(\tau/z,b)\left(1-z+2z^2\right)\;.
\eeq

\begin{figure}[ht]
\begin{center}
\epsfig{file=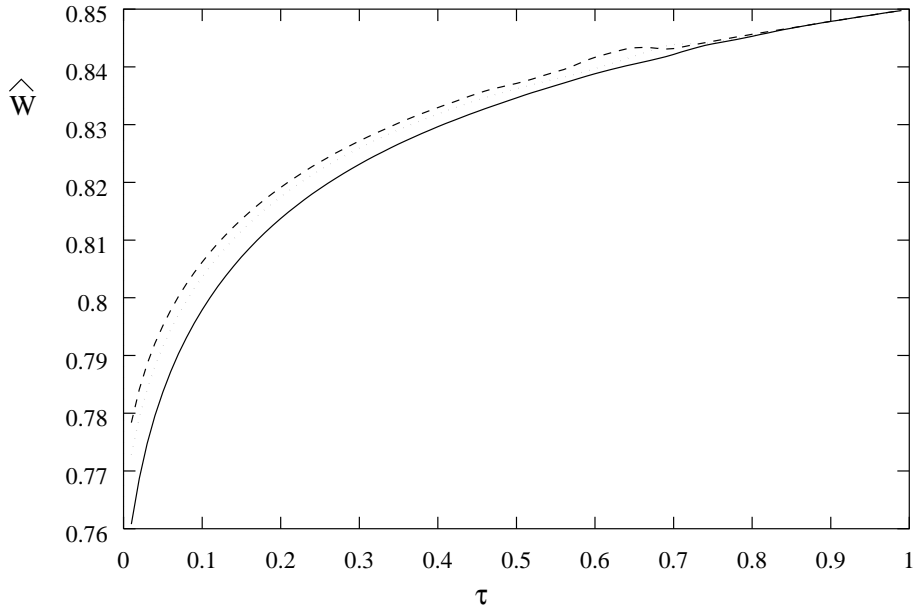}
\caption{\label{graph2}Graph of $\hat{W}(\tau,b)$ against $\tau$ for $b=0.5$ GeV$^{-1}$ (solid line), $b=0.1$ GeV$^{-1}$ (dotted) and $b=0.01$ GeV$^{-1}$ (dashed), evaluated at the Z mass with $\cl{A}_2=0.2$ GeV$^2$ and $\cl{A}_2^\prime=0$.}
\end{center}
\end{figure}

Figure \ref{graph2} shows a plot of $\hat{W}(\tau,b)$ against $\tau$ at fixed values of $b$. Again there is a slight logarithmic dependence of $\hat{W}(\tau,b)$ on $b$, coming from the scale of the parton distributions. Note however that $\hat{W}$ as a function of $\tau$ is very much flatter than $W$ was, since we have absorbed the most singular behaviour into the parton density functions. However this advantage is to some extent offset by the fact that the shift of about --3 GeV$^2$ in the factorisation scale reduces further the region of impact parameter space that may be treated perturbatively.

\section*{Acknowledgements}
We would like to thank the following for invaluable advice and helpful discussions: M.~Dasgupta, Yu.L.~Dokshitzer, G.~Marchesini, G.P.~Salam and B.R.~Webber. The work was supported in part by the EU Fourth Framework Programme `Training and Mobility of Researchers', Network `Quantum Chromodynamics and the Deep Structure of Elementary Particles', contract FMRX-CT98-0194 (DG 12 - MIHT).

\section*{Appendix}
\setcounter{equation}{0}
\renewcommand{\theequation}{A.\arabic{equation}}
In this appendix are included some technical details of the calculations.
\begin{enumerate}
\item Integral \rf{Creal}:

We require a series expansion in $\D$ for the integral
\beq
L(\D)=\int_0^{1+2\D-2\sqrt{\D(1+\D)}}dz\,z^{N-1}\frac{2+2z^2-4z\D}{\sqrt{(1-z)^2-4z\D}}\;.
\eeq
To this end we divide the integration region at $z=1-\sqrt{\D}/\k$, where $\sqrt{\D}\ll\k\ll 1$, and take each part separately. Since $\k$ is an arbitrary quantity, we consistently discard all terms that vanish as $\k\to0$.

For $0<z<1-\sqrt{\D}/\k$ we may expand the integrand directly in $\D$ to obtain
\bea
\label{apc1}
L_1(\D) &=& \int_0^{1-\sqrt{\D}/\k}dz\,z^{N-1}\left[\frac{2+2z^2}{1-z}+\cl{O}(\D)\right]\nln
&=& -2\log\frac{\D}{\k^2}-4S_1(N)-\frac{2}{N}-\frac{2}{N+1}+4N\frac{\sqrt{\D}}{\k}+\cl{O}(\D)\;,
\eea
where
\beq
S_i(N) = \sum_{n=1}^{N-1}\frac{1}{n^i}\;.
\eeq
The contribution from the second integration region can be evaluated by substituting $z=1+2\D-2\sqrt{\D(1+u+\D)}$ and expanding in $\D$ to obtain
\bea
\label{apc2}
L_2(\D) &=& \int_0^{\left(\frac{1}{2\k}+\sqrt{\D}\right)^2-1-\D}\frac{du}{\sqrt{u}}\left[\frac{2}{\sqrt{1+u}}-4N\sqrt{\D}+\cl{O}(\D)\right]\nln
&=& -2\log\k^2-4N\frac{\sqrt{\D}}{\k}+\cl{O}(\D)\;.
\eea
Adding together the two contributions \rf{apc1} and \rf{apc2} gives the infra-red divergent part as shown in \rf{apc3}. Note also that the term in $\sqrt{\D}$ cancels --- the fact that in the collinear approximation a $1/Q$ term remains indicates that it is not a valid approximation for studying power-suppressed corrections.

\newpage
\item Approximation \rf{besappr}:

The contributions $R_1$ and $R_2$ to the radiator are of the generic form
\beq
\label{radp}
R_i = \int_0^{Q^2}\frac{dk_\perp^2}{k_\perp^2}\left[1-J_0(bk_\perp)\right]f_i(k_\perp^2)\;.
\eeq
Let us define the function
\beq
\label{amu2q2}
F_i(\mu^2,Q^2) = -\int_{\mu^2}^{Q^2}\frac{dk_\perp^2}{k_\perp^2}f_i(k_\perp^2)\;.
\eeq
Then, integrating \rf{radp} by parts and making the substitution $x=bk_\perp$, we see that
\beq
\label{radimint}
R_i = \int_0^{bQ}dx\,F_i((x/b)^2,Q^2)J_0^\prime(x)\;.
\eeq
Further, by inserting the appropriate functional form of $f_i(k_\perp^2)$ into \rf{amu2q2}, we see that $F_i((x/b)^2,Q^2)$ is some function of $\log(x/b)$ which is slowly varying in $x$ compared with the oscillating $J_0^\prime(x)$. We may therefore extend the range of integration to infinity, and expand in $\log x$ about $x=b_0$:
\beq
F_i((x/b)^2,Q^2) = F_i((b_0/b)^2,Q^2)+F_i^\prime((b_0/b)^2,Q^2)\log(x/b_0)+\cdots\;,
\eeq
where here $F_i^\prime\equiv\partial F_i/\partial\log x$, and the higher derivatives have correspondingly fewer logarithms. Then, since
\beq
\int_0^\infty dx\,J_0^\prime(x) = -1 \ind
\int_0^\infty dx\,J_0^\prime(x)\log(x/b_0) = 0\;,
\eeq
we obtain from \rf{amu2q2}
\beq
R_i = \int_{(b_0/b)^2}^{Q^2}\frac{dk_\perp^2}{k_\perp^2}f_i(k_\perp^2)\;.
\eeq
Thus the Bessel function $1-J_0(bk_\perp)$ may be replaced in the radiator by the step function $\Theta(bk_\perp-b_0)$ to the accuracy we are using.

\item Integrals \rf{IJints}:

By substituting $u=\cosh\eta$, we may write
\bea
I_N(\e) = \int_1^\infty\frac{du}{u}\frac{\log(u+\sqrt{u^2-1})}{(1+\e+2\sqrt{\e}u)^N} &=& -\frac{1}{4\pi i}\int_C\frac{dz}{z}\frac{\log^2(z+\sqrt{z^2-1})}{(1+\e-2\sqrt{\e}z)^N}\nln
J_N(\e) = \int_1^\infty du\frac{\sqrt{u^2-1}}{(1+\e+2\sqrt{\e}u)^N} &=& -\frac{1}{2\pi i}\int_C dz\frac{\sqrt{z^2-1}\log(z+\sqrt{z^2-1})}{(1+\e-2\sqrt\e z)^N}\;,\hspace{1.0cm}
\eea
where the contour $C$ is:
\begin{center}
\epsfig{file=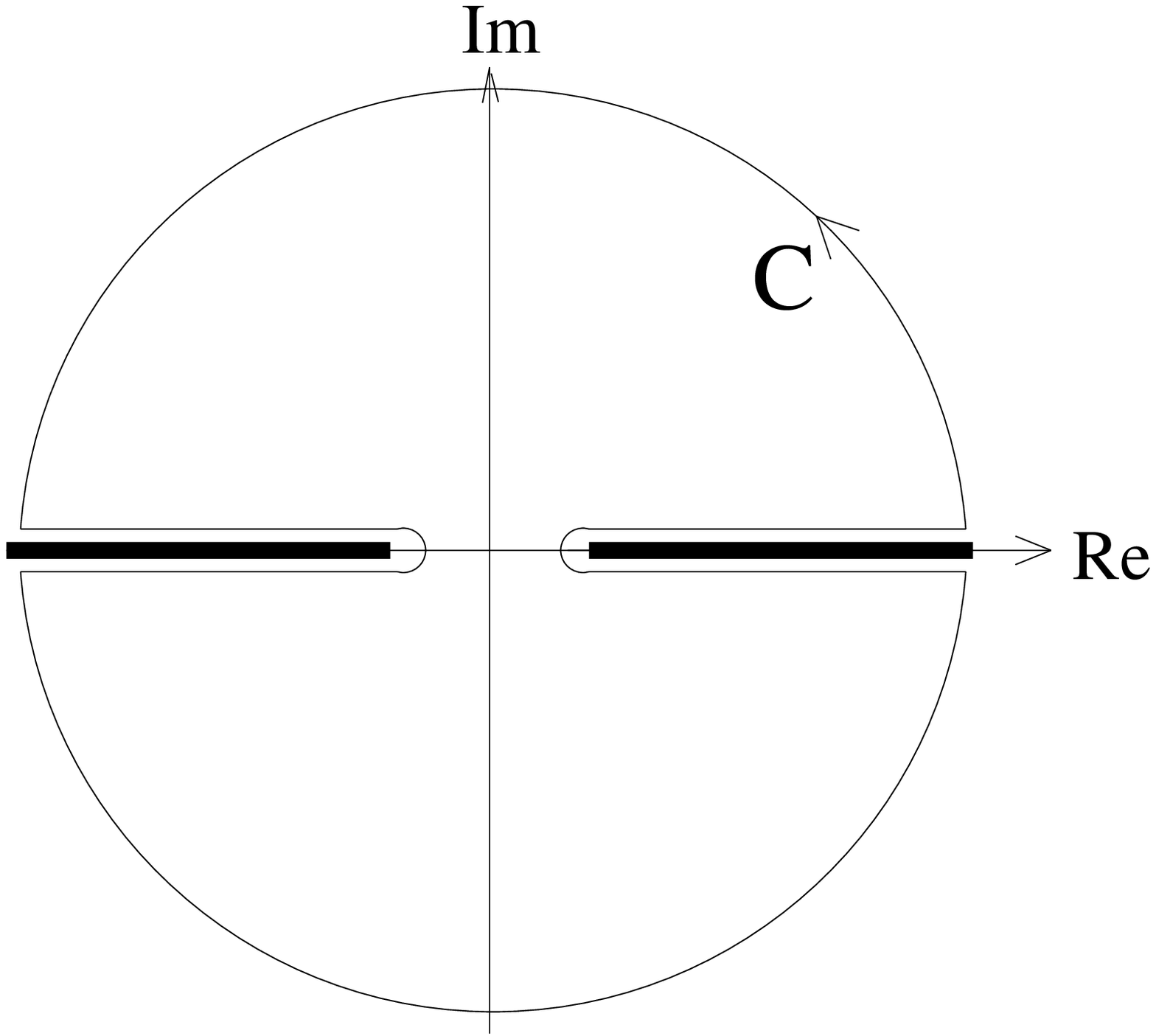, width=5cm}.
\end{center}
This contour encloses poles in the integrand at $z=0$ and $z=(1+\e)/(2\sqrt{\e})$. The residues of these poles may be evaluated by expanding around them, yielding the results:
\bea
I_N(\e) &=& \frac{1}{8}(1-N\e)\left[\pi^2-4S_2(N)+(2S_1(N)+\log\e)^2\right]\nln
& & \ind-\frac{\e}{4}\left[3N^2-N-2-(N-1)(N+2)(2S_1(N)+\log\e)\right]+\cl{O}(\e^2)\nln
J_N(\e) &=& \frac{1}{4\e}\left[\frac{1}{(N-1)(N-2)}+\e\left(2S_1(N)+\log\e-1-\frac{1}{N-1}\right)+\cl{O}(\e^2)\right]\;.
\eea
\end{enumerate}

\end{document}